\documentclass[sigconf]{acmart}
\setcopyright{none}
\acmDOI{}
\acmISBN{}
\acmConference{Preprint}{2026}{Online}
\acmBooktitle{Proceedings of Preprint}
\renewcommand{\footnotetextcopyrightpermission}[1]{}
\AtBeginDocument{%
  }

\usepackage{booktabs}
\usepackage{array}
\usepackage{multirow}
\usepackage{xcolor}
\usepackage{soul}
\begin{document}

\title{From Democracies to Autocracies: How AI Systems Enable Authoritarianism by Design}

\author{Jeba Sania}
\affiliation{%
  \institution{Harvard Kennedy School}
  \city{Cambridge, MA}
  \country{USA}
}
\email{jebasania@hks.harvard.edu}

\author{Marta Ziosi}
\affiliation{%
  \institution{University of Oxford}
  \city{Oxford}
  \country{UK}
}
\email{marta.ziosi@sant.ox.ac.uk}

\author{Fazl Barez}
\affiliation{%
  \institution{University of Oxford}
  \city{Oxford}
  \country{UK}
}
\email{fazl@robots.ox.ac.uk}

\renewcommand{\shortauthors}{Jeba Sania, Marta Ziosi \& Fazl Barez}

\begin{abstract}
AI-enabled authoritarianism is not confined to autocracies. In this paper, we provide greater transparency by investigating and mapping the lifecycles of six AI systems deployed in different political regimes, ranging from the US to China. By drawing on an extensive range of sources (academic publications, investigative research reports, third-party evaluations, media interviews, government procurement notices), we conduct a systematic, qualitative comparison across systems to identify the critical technical and operational features that enable authoritarianism within their respective political contexts. 

We find that enabling features include the centralization and co-optation of administrative data for law enforcement and political punishment, regulatory gaps that fail to deter misuse, weak user compliance that nullifies human oversight mechanisms, and the encoding of protected group traits that identify members of vulnerable populations. We find that these features are present across systems deployed in autocratic and democratic regimes, albeit in varying configurations. We also find that both centralized and fragmented AI systems can contribute to authoritarianism by exploiting governance gaps: centralized systems directed by executive authorities, particularly within security and military institutions, are often not subjected to formal oversight mechanisms, while fragmented systems diffuse accountability between stakeholders, paving the way for entrenchment.

These findings reveal that AI-enabled authoritarianism is distributed—resulting from design and operational choices made by developers, administrators, and users alike. We conclude with recommendations for developers and policymakers to mitigate these risks.
\end{abstract}



\maketitle

\section{Introduction}
The use of technology by anti-democratic regimes to maintain control is well-documented. Classic examples include Nazi Germany’s use of IBM technologies to streamline the logistics of facilitating genocide \cite{black_2001}, and the Chinese government’s efforts to control and censor the flow of information on the internet within the country\cite{roberts_2018}. With the rise of the internet, the phrase "digital authoritarianism" was coined to describe the use of “digital information technology by authoritarian regimes to surveil, repress, and manipulate domestic and foreign populations”\cite{erixon_lee-makiyama_2011,roberts_oosterom_2024,Pearson_2024,polyakova_meserole_2019}.

Artificial intelligence, through its rapid development pace, increasing capabilities, and wide adoption and application, expands the scope and negative impact of these technologies. New reports document how AI has been misused to destabilize democracies\cite{csernatoni_2024,polishchuk_2024}, interfere with elections\cite{csernatoni_2024,polishchuk_2024,unver_2024}, and manipulate at scale via misinformation in unforeseen ways globally \cite{csernatoni_2024,polishchuk_2024,unver_2024}. Key risks of AI technologies include highly accurate mass surveillance \cite{barez_friend_reid_krawczuk_wang_mökander_torr_morse_trager_2025}, reduced costs of monitoring and repressing dissent \cite{earl_maher_pan_2022,tokson_2025,beraja_kao_yang_yuchtman_2023}, and increased state capacity to manipulate public opinion\cite{funk_shahbaz_vesteinsson_2023}.

Despite this growing concern, systematic understanding of how these technologies are operated remains limited. Existing discussions primarily focus on system capabilities\cite{unver_2024,barez_friend_reid_krawczuk_wang_mökander_torr_morse_trager_2025,frantz_kendall-taylor_wright_2020} or downstream impacts\cite{hendrycks_authors_mazeika_2023,unver_hamid_akin_2018,bullock_hammond_krier_2025,mantellassi_2023}, \textbf{rather than how these AI systems are developed, deployed, and operated to advance authoritarianism.} This paper addresses this gap by shifting the focus from system capability to deployment by 1) investigating which features of AI systems (e.g model architecture, evaluation framework, launch plan) are key to enabling authoritarianism, and 2) moving beyond model-centric analysis by offering a systematic mapping of the lifecycle of the systems in question. In this paper, we distinguish between authoritarianism as a strict regime classification and authoritarianism as a set of practices that can be facilitated by technology across different political regimes.

We focus on the following AI systems: FlockSafety’s Automated License Plate Recognition (ALPR) System, Live Facial Recognition (LFR) Vans, SlimmeCheck, Lavender, Integrated Joint Operations Platform (IJOP), and Sfera. These systems were selected based on confirmation of the presence of AI, system type (e.g surveillance, predictive policing), political deployment context, and availability of public system documentation. We study these systems as deployed respectively in the United States, the United Kingdom, the Netherlands, the State of Israel, the People’s Republic of China, and the Russian Federation. 

Using a qualitative approach, we draw on investigative reporting, technical audits, system evaluations, and official government statements, among other sources. Through systematic cross-comparison of these systems across ten lifecycle stages, we identify the technical and contextual features that may enable authoritarianism. We examine whether these features have regime-specific patterns. Our approach emphasizes a socio-technical lens by treating system design and deployment as the primary objects of study. We also recognize that such an analysis inevitably intersects with political science to contextualize how these technical features manifest as authoritarian control. We call for future work, including our own, to engage in deeper interdisciplinary collaboration with experts in the field to enrich the analysis and balance overly technical perspectives.

As our main contribution, we provide a set of enabling factors through which system features advance the following characteristics of authoritarianism: expanded coercive capacity, erosion of accountability, symbolic safeguards, information control, anticipatory repression, and boundary control. We also find that while fragmented AI systems can contribute to authoritarianism like centralized systems, fragmented systems evade oversight mechanisms precisely due to their decentralized nature. Consequently, fragmented systems represent an underscrutinized mode of authoritarian risk. We then discuss the implications of these findings and provide takeaways for developers and policymakers. As this study relies on publicly available data of covert and controversial AI systems, there are inherent information gaps.

\section{Literature Review}

\subsection{Developing A Framework for Understanding Authoritarianism}

\subsubsection{Defining Authoritarianism}
Authoritarianism is commonly defined as a political system with limited political pluralism, competition, or accountable executive authority\cite{linz_1964,schedler_2002}. While this literature conceptualizes authoritarianism primarily as a regime type, this study adopts a practice-oriented perspective, focusing on authoritarian mechanisms and practices that may occur both within and beyond formally authoritarian regimes. Authoritarian rulers often minimize constraints on their power while repressing or strategically containing meaningful political contestation \cite{levitsky_way_2002,glasius_2018,brown_schaaf_anabtawi_waller_2024}. The mechanisms that enable authoritarianism are highly contextual and vary across different regions and time periods \cite{slater_2018,libman_obydenkova_2018,velez_lavine_2017,ekiert_2023,owen_2022}.

Despite contextual variations, scholars have identified recurring features of authoritarianism. This section reviews existing literature to set up a conceptual framework outlining six interrelated characteristics of authoritarian systems. 
\subsubsection{Foundational Frameworks: Centralization and Control}
Foundational works emphasize the centralization of power as a defining element of authoritarianism. Early theorists Friedrich and Brzezinski demonstrate that high degrees of power centralization occur across different spheres, including information channels and means of force \cite{friedrich_brzezinski_1965}. Multiple scholars emphasize the state’s capacity to centralize means of coercion \cite{friedrich_brzezinski_1965,wintrobe_1998,davenport_2007,slater_2010}. Scott's analysis of state legibility in "Seeing Like a State" provides a foundational account of how states, more generally, have sought to centralize information as a precondition for administrative control\cite{Scott_1998}. As an example of an extreme form of coercion, Mbembe's concept of necropolitics explores the state's power to dictate who may live and who must die, which may entail terminating as well as ignoring, neglecting, or surveilling the lives of marginalized groups\cite{Mbembe_Corcoran_2019}. Across this literature, coercion is understood not only as violent repression but as a system of threats and intimidation to incentivize certain political outcomes. These works highlight the state’s ability to enforce compliance by monopolizing the use or threat of violence to achieve political goals, rather than democratic means.

\subsubsection{The Role of Institutions: From Obstacles to Tools}
Another mechanism authoritarian governments employ to centralize power is by undermining institutions, such as the electorate, judiciary, and the media\cite{friedrich_brzezinski_1965}. Linz, in his seminal 1964 "An Authoritarian Regime: The Case of Spain," includes “ill-defined executive powers that routinely surpass democratic limits” as part of his three-part definition of authoritarianism\cite{linz_1964}. Overreaching executive powers undermine other forms of institutional authority meant to serve as checks. 

Various scholars maintain an alternative view: rather than eroding institutions, authoritarian regimes strategically employ institutions to consolidate power\cite{gandhi_2008,frantz_kendall-taylor_2014,svolik_2012}. Institutions are not obstacles, but rather tools to manage the key tension of appeasing ruling elites and repressing citizens\cite{gandhi_2008,svolik_2012}. According to this view, institutions merely provide symbolic accountability. After synthesizing decades of comparative authoritarianism literature, Frantz and Kendall-Taylor, in their "A Dictator's Toolkit: Understanding How Co-optation, Repression, and Legitimization Help Autocrats Survive," identify legitimation as a key authoritarian strategy to build positive public perception\cite{frantz_kendall-taylor_2014}. Collectively, the existing scholarship suggests that authoritarian rulers counter existing accountability mechanisms, either directly or indirectly.
\subsubsection{Enduring Authoritarianism: Subtle Manipulation Over Overt Repression}
Contemporary scholarship highlights subtler mechanisms through which modern authoritarian regimes maintain power. Schedler describes a “menu of manipulation” where regimes structure law enforcement and political competition to preserve dominance while projecting a façade of legality\cite{schedler_2015,schedler_2002}. Rather than relying on outright repression, authoritarians often manipulate information as a low-cost alternative\cite{guriev_treisman_2020}. Propaganda and censorship are tools to manage public opinion and maintain the appearance of democracy, thus preserving legitimacy in the international arena\cite{treisman_guriev_2023}.

At the same time, the literature underscores the role of pre-emptive modes of repression, such as harassment and threats to deter wider opposition \cite{lorentzen_2014,king_pan_roberts_2013}. Frantz and Kendall-Taylor note that anticipatory tactics, such as surveillance or intimidation, complement coercive approaches by shaping societal expectations of political punishment\cite{frantz_kendall-taylor_2014}. Foucault's analysis of the panopticon provides a useful lens for understanding how anticipatory punishment, surveillance-induced compliance, and chilling effects enforce the disciplinary power of institutions\cite{Foucault_1977}. In summary, the longevity of authoritarianism depends on strategic combinations of subtler mechanisms with targeted, pre-emptive forms of repression.

\subsubsection{A Persistent Feature: Insider/Outsider Boundaries}
Another recurrent feature is the exclusion of minority populations. Linz highlights limited pluralism in his definition of authoritarianism, a condition that sets the stage for systemic social, political, and economic discriminatory boundaries\cite{linz_1964}. Arendt's foundational analysis of Nazi Germany and Stalinist Russia examines how specific populations were excluded from state-granted legal protections, civic rights, and human rights, effectively denying what she terms the “right to have rights”\cite{arendt_1968}. While Arendt’s analysis applies to extreme cases, later work shows that exclusionary boundaries are common across authoritarian contexts\cite{cheeseman_2015,king_pan_roberts_2013,brownlee_2007}. Other scholars provide nuance by conceding that exclusionary politics can paradoxically also put authoritarian regimes at risk of mass opposition \cite{cederman_wimmer_min_2010}. However, strategies such as power-sharing with elites and balancing ruling coalitions mitigate these risks\cite{svolik_2012,beiser-mcgrath_metternich_2020}. Postcolonial scholarship reveals that the specific mechanisms enforcing such boundaries have deep institutional histories, as evidenced by Breckenridge's account of biometric governance in colonial India as a tool of racial classification\cite{Breckenridge_2016} and Fanon's analysis of colonial population division\cite{Fanon_Sartre_1963}.  Collectively, these works illustrate how erosion of the "right to have rights" underpins exclusionary 'inside/outsider' classification.

\subsubsection{Synthesizing A Conceptual Framework of Authoritarianism}
Drawing from the literature, the following conceptual framework synthesizes key characteristics of authoritarian governance. Together, these characteristics provide an analytical framework to contextualize the six AI-enabled authoritarian deployments in the following sections.
\begin{itemize}
\item \textit{Coercive Capacity} is defined as the extent to which force and other forms of compulsion are monopolized by executive powers.
\item \textit{Accountability Erosion} is defined as the removal or weakening of formal or informal oversight mechanisms.
\item \textit{Symbolic Safeguards} are defined as accountability processes without genuine or meaningful enforcement that can be exploited or co-opted, and thus become largely symbolic. 
\item \textit{Information Control} is defined as the monopolization and manipulation of information channels, including invasive monitoring and surveillance practices. 
\item \textit{Anticipatory Repression} is defined as pre-emptive acts of coercion, intimidation, and harassment to maintain control.
\item \textit{Boundary Control} is defined as drawing exclusionary social and political categories that violate the political participation and human rights of specific populations, particularly minorities.
\end{itemize}
It is important to caveat that these characteristics are not strictly confined to the authoritarian-democratic spectrum of governance, e.g., colonial administrative states as discussed previously\cite{Breckenridge_2016,Scott_1998,Chatterjee_1993}. They are also not fully exhaustive, but they aim to capture salient characteristics that enable authoritarian outcomes.

\subsection{AI Technologies as Enablers of Authoritarianism}

Authoritarians have historically resorted to technology to enforce their rule \cite{friedrich_brzezinski_1965,lamensch_2021}. Often referred to as digital authoritarianism or techno-authoritarianism, this field studies how authoritarianism evolves as technological systems’ capabilities increase \cite{mantellassi_2023,mentxaka_díaz-rodríguez_coeckelbergh_lópez_gómez_llorca_herrera-viedma_herrera_2025,peron_almstadter_mattar_de_magalhães_caetano_2025,schlumberger_edel_maati_koray_saglam_2023}. While these terms have slight definitional variations, they describe the use of technology to enable authoritarian practices, irrespective of formal regime classification. This reflects how AI systems can facilitate the adoption of authoritarian practices in multiple contexts.
 For the purposes of this study, we define AI systems, following the OECD, as machine-based systems that generate predictions, recommendations, or decisions influencing physical or virtual environments\cite{oecd_2024}. Such systems are of particular interest due to specific features such as statistical learning, extensive data collection and analysis, and black-box algorithmic decision-making that resists explainability and thus evades effective oversight \cite{barez_friend_reid_krawczuk_wang_mökander_torr_morse_trager_2025,pastor-galindo_nespoli_ruipérez-valiente_2024,sofia_anastasiadou_santos_2024}. While AI systems do not originate authoritarian practices, nor do they uniquely produce the characteristics of authoritarian governance as presented in Section 2.1.6, they can exacerbate such characteristics, extending reach, reducing cost, and obscuring accountability\cite{beraja_kao_yang_yuchtman_2023}. There are several key domains in which AI enables authoritarianism, including mass surveillance, predictive policing, and the manipulation or control of information systems \cite{barez_friend_reid_krawczuk_wang_mökander_torr_morse_trager_2025}.

\subsubsection{Mass Surveillance: Monitoring at Unprecedented Scale}
Digital surveillance systems expand state capacity to monitor citizens at an unprecedented scale by automating mass data collection and predictive identification via machine learning models. While state surveillance predates AI, AI systems introduce a dramatic reduction in the cost and labor required to sustain surveillance at scale\cite{beraja_kao_yang_yuchtman_2023}. This creates chilling effects and violates citizens’ fundamental right to privacy and free speech \cite{vagianos_stavrou_2023,neary_2022}. To support such endeavors, regimes invest in surveillance infrastructure such as advanced hardware, highly integrated datasets, and biometric models. Examples of mass surveillance AI systems include China's Social Credit System, which aggregates and classifies large amounts of data for behavioral scoring\cite{liang_das_kostyuk_hussain_2018}.
\subsubsection{Predictive Policing: Pre-emptive Control over Opposition}
Surveillance systems feed into predictive policing systems that enable anticipatory repression. Predictive analytics identify dissenters and preempt opposition before mobilization. In practice, ill-defined thresholds of suspicion include civilian protestors or even entire ethnic or religious minorities perceived to be in opposition to the current regime \cite{lau_2020}.These technologies contribute to self-censorship, limited political engagement, and reduced pluralism \cite{strikwerda_2020}. Multiple police departments throughout the United States, in partnership with private sector companies have deployed predictive policing systems\cite{ziosi_pruss_2024,richardson_schultz_crawford_2019}.
\subsubsection{Information Manipulation: Controlling Narratives and Public Discourse}
Information-manipulation technologies disseminate propaganda and suppress ideas critical to the ruling regime \cite{chen_xu_2015,guriev_treisman_2020}. Limited transparent, free information biases public discourse, challenging a society’s ability to align on shared facts and solve problems \cite{nie_2024,ajayi_2025}. Examples include AI-driven disinformation campaigns and content moderation algorithms that censor at scale \cite{miner_natsika_lindberg_2024}. Non-AI examples include control over digital infrastructure such as internet shutdowns, and bans on news and social media platforms\cite{miner_natsika_lindberg_2024}.
\subsubsection{Global Proliferation}
Authoritarian practices, such as the expansion of surveillance infrastructure, are not strictly confined to historically authoritarian regimes. This is evidenced by the international market for surveillance technologies \cite{zuboff_2019,ni_2021,schuppe_2019,davies_abraham_2025}. Digital authoritarianism is spreading globally, even in liberal democracies such as the US and those in Western Europe \cite{richardson_schultz_crawford_2019}. The lack of oversight, enforceable accountability, and safeguards contributes to this rise, intentionally and unintentionally, domestically and internationally \cite{mentxaka_díaz-rodríguez_coeckelbergh_lópez_gómez_llorca_herrera-viedma_herrera_2025}.

\section{Research Question}

Our driving research question is, \textit{“Which lifecycle features and contextual factors act as key enablers to facilitate authoritarian practices and governance outcomes, and do these factors vary across regimes?}” We hypothesize that these factors will vary significantly between the liberal democratic, authoritarian, and hybrid regimes. 

\section{Methodology}
To examine the deployment of authoritarian AI systems, we employed a case study approach\cite{gomm_hammersley_foster_2009}, mapping out the lifecycle of six systems applied in different political regimes across ten dimensions outlined in Section 4.1. To ensure consistency and comparability, each system was analyzed against these same dimensions and the conceptual framework established in Section 2. We then conducted a comparative analysis to identify patterns\cite{mills_rihoux_2008}, while remaining sensitive to the varying political and operational contexts of each deployment.
\subsection{Dimension Selection}
To provide a clear, structural understanding of systems’ deployment lifecycles, we selected dimensions adapted from conventional product development frameworks \cite{zhang_subramani,raeburn_2024}. These dimensions capture critical technical and non-technical stages, and enable systematic\footnote{We use the term 'systematic' here to denote a structured, consistent analytical approach applied uniformly across all six cases. Specifically, we refer to the comparison of each case against the same ten lifecycle dimensions rather than a specific review protocol.} comparison across cases.

The ten lifecycle dimensions are as follows: 
\begin{itemize}
\item \textit{Ideation}: Which events or political drivers motivated its development, who defined the system’s goal, and which stakeholders participated and gave approval for its initial conception? 
\item \textit{Mandate and Legitimization}: How is the system’s purpose justified, framed, and/or communicated to the public? This includes examining official announcements, media narratives, and public reception.  
\item \textit{Procurement and Partnership}: How are the system’s components, technologies, and talent to operate the system acquired? Which procurement processes, funding arrangements, and public-private partnerships were utilized? 
\item \textit{Design and Development}: Which stakeholders drove the technical development process? This dimension also considers whether the system is an iteration of a previous system, and the technical stack used, including data sources, compute requirements, and model specifications.
\item \textit{Technical Integration}:  To what extent is the system integrated with other technical tools, especially existing government systems? 
\item \textit{Testing}: How was the system’s evaluation designed? How did the system perform, and which stakeholders conducted the evaluation? 
\item \textit{Operational Rollout}: How was the system launched, and was its rollout communicated? Which stakeholder is responsible for its ongoing maintenance, and what is the extent of public controversy surrounding its launch? 
\item \textit{Oversight}: What governance mechanisms are in place to monitor the system, such as audits, legal challenges, and third-party reviews? How independent are those mechanisms?
\item \textit{Safeguards}: Are there technical and procedural protections built into the system itself, such as data retention limits and audit logs? This dimension also considers whether any safeguards were overridden.  
\item \textit{Current State}: How has the system evolved (or not) since its initial deployment as of the time of writing, and which stakeholders exert influence over the system’s direction? 
\end{itemize}
\subsection{System Selection}
The criteria for which these AI systems were chosen include:
\begin{itemize}
\item \textit{Presence of Artificial Intelligence}: Is there credible evidence that these systems employ machine learning or other advanced statistical approaches? 
\item \textit{System Classification}: Can these systems be classified as surveillance, predictive policing, or information control systems as defined by Barez et al.'s framework of AI-enabled authoritarian systems \cite{barez_friend_reid_krawczuk_wang_mökander_torr_morse_trager_2025}?
\item \textit{Differing Political Contexts}: Is there variation in the political regimes that these systems originate from and are deployed in? To assess this, we draw on well-respected nonprofit democracy indexes\footnote{ We recognize that these  classifications have certain normative assumptions and inherently reflect historical and political contexts. In this study, we use these classifications primarily to contextualize the political environments these systems are deployed in.
}, including the V-Dem Institute’s \textit{Democracy Report} and Freedom House’s \textit{Democracy In The World}.
\item \textit{Deployment Data Availability}: Is there sufficient public information on these systems available and in the English language (the working language of the author) that verifies the presence of these systems and their lifecycles? 
\end{itemize}
The systems are presented in Table 1. Further information on how each system meets the criteria outlined above can be found in Appendix Table 2.

Despite differences in scale, institutional purpose, and political context between the six systems, we treat them as analytically equivalent units of comparison on a specific and limited basis: AI systems whose development and deployment can be mapped across the ten lifecycle dimensions in Section 4.1. This framing allows for the comparison of diverse systems on the basis of high-level categories of design and deployment features and the evidence for them, rather than scale, purpose, or impact. 
\begin{table*}[t]
\centering
\caption{Overview of selected AI systems.}
\label{tab:system_intro}
\begin{tabular}{%
  p{0.15\textwidth}
  p{0.15\textwidth}
  p{0.12\textwidth}
  p{0.15\textwidth}
  p{0.30\textwidth}
}
\toprule
\textbf{System Name} & \textbf{Deployment Country} & \textbf{Domain} &
\textbf{Political Regime Classification} & \textbf{System Description} \\
\midrule
FlockSafety Automated License Plate Recognition (ALPR) &
United States &
Policing &
Liberal Democracy\cite{nord_altman_angiolillo_fernandes_good_god_lindberg_2024}, Free\cite{gorokhovskaia_grothe_2025} &
Neighborhood and business-level license plate recognition camera system developed by a private US company, FlockSafety\cite{flocksafety_2025c}. \\

Live Facial Recognition (LFR) Vans &
United Kingdom &
Policing &
Electoral Democracy\cite{nord_altman_angiolillo_fernandes_good_god_lindberg_2024}, Free\cite{gorokhovskaia_grothe_2025} &
Real-time facial recognition vans used by 13 local UK police departments for suspect identification\cite{office_2025b}. \\

SlimmeCheck &
The Netherlands, Amsterdam &
Administrative (Welfare Fraud) &
Liberal Democracy\cite{nord_altman_angiolillo_fernandes_good_god_lindberg_2024}, Free\cite{gorokhovskaia_grothe_2025} &
Welfare fraud detection system developed by the Amsterdam municipality government’s Work, Participation, and Income (WPI) department\cite{guo_geiger_braun_2025}. \\

Lavender &
Israel, Gaza Strip &
Military &
Electoral Democracy\cite{nord_altman_angiolillo_fernandes_good_god_lindberg_2024}, Free\cite{gorokhovskaia_grothe_2025} &
Militant identification system developed by the Israeli Defense Force’s Unit 8200\cite{abraham_2024a}. \\

Integrated Joint Operations Platform (IJOP) &
China, Xinjiang Province &
Policing &
Closed Autocracy\cite{nord_altman_angiolillo_fernandes_good_god_lindberg_2024}, Not Free\cite{gorokhovskaia_grothe_2025} &
Aggregated policing system used to monitor and identify security threats in Xinjiang, developed by the Xinjiang Public Security Bureau\cite{wang_2019}. \\

Sfera &
Russia, Moscow &
Transportation &
Electoral Autocracy\cite{nord_altman_angiolillo_fernandes_good_god_lindberg_2024}, Not Free\cite{gorokhovskaia_grothe_2025} &
Biometric (facial recognition) payment system used in the Moscow Metro developed by Russian private-public partnerships\cite{ovd-info_2022}. \\
\bottomrule
\end{tabular}
\end{table*}

\subsection{Case Study Methodology}

We drew on a wide range of sources to ensure the comprehensiveness of our analysis. For each system, we conducted searches combining system names with terms related to lifecycle dimensions. We routinely followed citations in identified sources to locate additional material. While the availability of information varied across systems, we consulted the same categories of sources for each system to minimize information asymmetries in our comparative analysis.

These source categories include investigative research reports, human rights reports and investigations from well-known NGOs, first-hand accounts from impacted individuals of the public, government employees, and system developers, public procurement notices, technical audits, independent system evaluations, informal statements by government officials via social media, news and media coverage, official government communications, public lectures and publications by system developers, tribunal testimonies, official technical system diagrams, leaked screenshots of software interfaces, and interviews with surveillance experts. Defunct links were cross-checked against the Internet Archive’s Wayback Machine. 

The diversity of sources enabled us to maximize the amount of credible information available on these systems. To maintain information fidelity, we excluded the following sources: opinion blogs, unverified social media posts, and materials that did not provide sufficient evidence to substantiate claims. 

The majority of sources were primarily examined over a one-month period from July to August 2025. Using the information gathered, we mapped each system against the ten lifecycle dimensions in Section 4.1, identifying technical and operational features based on whether they enabled the six authoritarian characteristics established in the Conceptual Framework in Section 2.1.6.

As an initial attempt to examine AI-enabled authoritarianism across regime types, this study prioritizes identifying the authoritarian-enabling features before evaluating the countervailing features that might mitigate authoritarian outcomes. We treat the latter as an important but distinct analytical question, which we leave for future work.

\section{Results}

Across AI systems, we identify technical and contextual features that enable authoritarian practices. Rather than organizing findings by political regime, country, or AI system, we structure this section primarily around the authoritarian characteristics these features enable and the mechanisms by which they do so, in Section 5.1. See Appendix Table 3 for a visual mapping. In Section 5.2, we report other notable results, including the advantages of centralized and fragmented systems in enabling authoritarianism, as well as findings on the limited transparency of these systems. We find patterns across systems, regardless of the political regime in which these systems are deployed, suggesting regime classification does not fully account for the presence of enabling features. 

\subsection{System Features}
\subsubsection{Coercive Capacity}
Coercive capacity is defined as the extent to which force and other forms of compulsion (harassment, intimidation, etc) are monopolized by an executive power. We also consider the extent to which threats of force, force, and its derivatives are made viable. 

In Gaza, where the Lavender system has been used to identify and recommend military targets, coercive capacity is expanded by a combination of three factors: integration across military systems, loose definitions of what constitutes a military target, and a 10\% identification error margin\cite{abraham_2024,abraham_2024a}. Collectively, these factors lower the threshold for lethal force. Lavender uniquely enables fatal outcomes among other systems. 

\textit{Feature: Integration with Lethal Military Infrastructure, Insufficient Accuracy Rates}

Across the US, FlockSafety’s deployment of local ALPR cameras in over 6,000 cities, along with the integration of state and national databases such as the Federal Bureau of Investigation’s National Crime Information Center (NCIC) database, significantly expands surveillance infrastructure\cite{flocksafety_2025}. Such expansion enables greater levels of harassment and the implicit threat of it. Notably, many national law enforcement databases have been reported to be routinely out of date or erroneous, increasing the risk of unwarranted harassment\cite{neighly_emsellem_2013,hand_1982,laudon_1986}. 

\textit{Feature: Widespread Data Integration, Integration with Outdated Databases}

Sfera, LFR Vans, and the IJOP system in particular, follow a similar model of centralizing surveillance inputs and wide integration with other databases \cite{fussey_murray_2019,wang_2019}. The Sfera system leverages data from non-law enforcement databases, including public services, local administration, and even passport databases \cite{ovd-info_2022,kommersant_2021}. It's worth mentioning that Sfera’s facial recognition models are partially trained on a dataset extracted from a doxxing mobile application, explicitly connecting surveillance capabilities to harassment functions \cite{signorelli_2024}. 

\textit{Feature: Widespread Data Integration, Co-optation of Administrative systems}
\subsubsection{Accountability Erosion}
Accountability erosion is defined as the removal or weakening of formal or informal oversight mechanisms. This dilution or denial of checks and balances is most relevant for the concentration of executive power. 

In the UK, police operators of LFR vans have wide discretionary power. Third-party auditors at the University of Essex note the freedom for operators to determine the proportionate use of the technology, suggesting limited constraints on improper usage\cite{whannel_2025,fussey_murray_2019,essex_2019}. Additionally, inconsistent operating procedures between deployments can diffuse responsibility for negative impacts and enable potential misuse \cite{whannel_2025,fussey_murray_2019,essex_2019}. This is made possible by the lack of regulation of facial recognition technology in the UK, which instead relies on a patchwork of recommendations\cite{oxley_uwazuruike_lalic_samuel_downs_2024,gov.uk_2013}.

\textit{Feature: Inconsistent Operating Procedures
}
The extent of automation within the Lavender system provides operators the option to defer decision-making to the system, which consequently limits opportunities for meaningful contestation \cite{alon-barkat_busuioc_2023,abraham_2024a}. 

\textit{Feature: Over-reliance on Automated Decision-Making
}
As the customers of FlockSafety’s ALPR system, businesses, HOAs, and private citizens own all locally collected surveillance data\cite{flocksafety_2025b}. However, these non-state users do not receive legal or ethical training and generally enjoy weaker oversight standards \cite{dean_2019,stanley_2022}. In a similar vein, informal data-sharing agreements between private citizens and law enforcement officials provide federal agencies with workarounds to avoid breaking state laws on ALPR data \cite{mohamed_2025,koebler_2025,mount_prospect_police_department_2023} and legal constraints in federal contracts \cite{maass_alajaji_2025,cox_koebler_2025,cushing_2024}. 

\textit{Feature: Oversight Loopholes for Non-state Actors and Private Citizens/Organizations}

Russia’s 2022 law on biometric data stipulates that its restrictions on the handling, retention, and transfer of data do not apply to intelligence, national security, or other law enforcement activities, creating a significant accountability gap\cite{grigoryev_2024,startsev_2022}. 

\textit{Feature: Regulatory Gap and Exemption for Law Enforcement 
}
\subsubsection{Symbolic Safeguards}
Symbolic safeguards are defined as existing accountability processes that lack genuine or meaningful enforcement, allowing for co-optation. Unlike accountability erosion, symbolic safeguards are maintained, but their presence serves to legitimize deployments rather than constrain.  

An official statement by the Israeli Defense Force justifies the usage of Lavender, indicating that analysts review and verify whether targets meet the criteria stipulated by international law and internal directives\cite{idf_2024}. However, investigative reports indicate that analysts intentionally bypassed such safeguards, spending only 20 seconds to verify whether identified targets were males in order to meet performance standards\cite{abraham_2024a}.  

\textit{Feature: Lack of Compliance Mechanisms 
}

Similarly, law enforcement users with access to FlockSafety’s ALPR surveillance data are required to log their purpose of searching the database. However, investigations have shown that queries often use vague language, rendering audits of the usage of this system ineffective \cite{maass_alajaji_2025,mohamed_2025}.

\textit{Feature: Inconsistent Operating Procedures, Lack of Compliance Mechanisms
}\subsubsection{Information Control}
Information control is defined as the monopolization and manipulation of information channels, including invasive monitoring and surveillance practices. Several of the examined systems consolidate large amounts of personal data. 

This is most evident through the IJOP system, deployed in Xinjiang. Government officials centralize surveillance data on the Uyghur ethnic minority population through extensive biometric, medical, utility, location, and digital communications data collection across public and private spaces \cite{peterson_2021,wang_2019b,oztig_celil_karluk_2025,lacin_idil_oztig_2023,wang_2019}.  Information from non-law enforcement databases is also connected to the IJOP system, such as second-hand car databases \cite{department_of_commerce_of_xinjiang_uygur_autonomous_region_2019}. 

\textit{Feature: Widespread Data Integration, Co-optation of Administrative Systems
}
While the IJOP system is justified on security grounds\cite{human_rights_watch_2018}, the Sfera system was introduced as a modernization effort to streamline payments for riders on the Moscow metro \cite{sobyanin_2026,pjotr_sauer_2021}. However, reports show that as of 2022, 141 Moscow residents identified through Sfera were effectively detained for participation in protests \cite{ovd-info_2022a}. The expansion of Sfera’s scope to target political dissenters exemplifies how executive powers manipulate information channels for political control.

\textit{Feature: Co-optation of Administrative Systems 
}
\subsubsection{Anticipatory Repression}
Anticipatory repression is a pre-emptive act of coercion, intimidation, and harassment without due process. Systems can enable anticipatory repression by identifying perceived political risks ahead of the fact.

Many residents who have participated in past anti-government protests in Moscow reported being arrested by police ahead of large holidays \cite{masri_2023,ovd-info_2022a}. These arrests occurred prior to any planned protest activity, illustrating that Sfera was used to prevent political mobilization.

\textit{Feature: Co-optation of Administrative Systems
}

Similarly, in Xinjiang, the IJOP system operationalizes anticipatory repression through the use of haphazard and arbitrary identifiers of criminality, such as emotions \cite{wakefield_2021,ipvm_2021}. Further arbitrary measures include changes in facial expressions and skin pores \cite{wakefield_2021}. Additionally, despite the advanced computer vision techniques used to collect inputs, the system’s downstream decision-making is comparatively unsophisticated. Investigative reports reveal that simple rules-based logic is used to determine detentions rather than ML-based predictions \cite{arelemorgen_2018,wang_2019}. This combination of advanced surveillance methods and lack of accurate decision logic enables mass detention. 

\textit{Feature: Predictive Behavioral Indicators, Arbitrary Indicators}

\subsubsection{Boundary Control}
Boundary control is the creation of exclusionary social and political categories that disadvantage specific populations, particularly minority groups. 

The IJOP system explicitly targets ethnic Uyghurs for detention and re-education camps in Xinjiang. This is evidenced by multiple “Uyghur detection” algorithms connected to the surveillance cameras feeding into the system \cite{healy_2024,ipvm_editorial_staff_2021,byler_2020,kelion_2021,rollet_2019}. Additionally, officials can report the ethnic and religious practices of residents in IJOP, including donating to mosques or preaching without authorization \cite{wang_2019,aspi_2018}. 

\textit{Feature: Explicit Group-Based Classification
}

In Gaza, individuals are more likely to be identified  by the Lavender system as a military target according to imprecise and somewhat arbitrary behaviors, including changing phones and addresses every few months, as is common during war \cite{clarno_2024,y.s_2021}. The lack of tight definitions can result in indiscriminate violence against Gazans.

\textit{Feature: Arbitrary Indicators }

During the testing and pre-pilot of SlimmeCheck, developers found that the algorithm was more likely to incorrectly flag applications from non-Dutch citizens and those with non-Western nationalities \cite{braun_geiger_guo_constantaras_silverman_2025}. 

\textit{Feature: Encoding of Protected Group Characteristics}

\subsection{Other Notable Patterns and Findings}
We refer to system centralization as the extent to which executive authorities meaningfully control or influence the AI system, as well as how connected the system is to multiple inputs and data sources. In line with this definition, the IJOP and Sfera systems are the most centralized, whereas Flock Safety’s ALPR system is the most fragmented, as it is operated by a private entity and exists as a collection of deployments.  

By concentrating decision-making authority, centralized systems are structurally well-suited to authoritarian applications. These systems have distinct authoritarian advantages stemming from the reduced internal and legal checks. This includes fewer institutional stakeholders with power to veto or constrain the system. Consequently, such systems are susceptible to oversight capture, mitigating the effectiveness of formal oversight mechanisms. For example, the Sfera system is exempt from any legal oversight for biometric systems in Russia, while operators of the Lavender system did not verify the system’s outputs against the IDF’s own definitions of military targets under international humanitarian law \cite{abraham_2024a}. In theory, centralization also facilitates an easier pathway to further integration with other government data repositories \cite{rujano_jan-willem_boiten_ohmann_canham_contrino_david_ewbank_filippone_connellan_custers_et_al._2024,kernstock_harms_hein_krcmar_2025}.

However, fragmented systems also pose unique authoritarian-enabling risks. Using FlockSafety’s ALPR as a case study, we note that attempts to regulate the technology have proven challenging. The decentralized manner of the system and its network effects make it harder to be collectively dismantled in the absence of centralized decision-making \cite{stanley_2022}. The ALPR network exhibits a lack of strong user coordination between deployments, resulting in a lack of clear responsible actors between FlockSafety, law enforcement users, and customers (neighborhoods, HOAs, and businesses). This is significant because Flock Safety lacks the ability to detect user-initiated abuse of the system \cite{dean_2019,hamid_alajaji_2025}. Lastly, inconsistent user-led deployments can lead to inconsistent implementation of best privacy practices and safeguards. 

Across the ten lifecycle stages, several patterns emerge across systems, emphasizing a lack of clean distinction between system features and regime classifications. At the mandate and legitimization stage, all systems, except for SlimmeCheck and Sfera, were justified explicitly through public safety and law enforcement rationales \cite{human_rights_watch_2018,tauvod_2023,sobyanin_2026,flocksafety_2025c,guo_geiger_braun_2025,corp_2025}. Importantly, the role of a preceding security crisis was a salient factor for most of these systems (IJOP, FlockSafety ALPR, Lavender )\cite{y.s_2021,bbc_2014,amnesty_international_2012,langley_2025}. At the procurement and partnership stage, most systems were either procured from private vendors or developed and fine-tuned through private-public partnerships\cite{wang_2019,flocksafety_2025c,south_wales_police_2017,nec_2017,coldewey_2016,masri_2023}. SlimmeCheck and Lavender were instead developed in-house by government institutions\cite{human_rights_watch_2024,tauvod_2023,yaniv_avital_2023,guo_geiger_braun_2025}. At the operational rollout stage, all systems except for FlockSafety ALPR deployments and the Lavender system are documented to be launched through a pilot approach \cite{oxley_uwazuruike_lalic_samuel_downs_2024,baraniuk_2017,south_wales_police_2017,braun_geiger_guo_constantaras_silverman_2025,oster_2016,moscow_metro_2022}. However, pilot results have been disclosed for SlimmeCheck, LFR Vans, and Sfera, but not for the IJOP system \cite{bbc_news_2018,braun_geiger_guo_constantaras_silverman_2025,moscow_metro_2022}. Regarding the current stage of systems, only SlimmeCheck was formally dismantled by city officials after a series of bias audits\cite{braun_geiger_guo_constantaras_silverman_2025,guo_geiger_braun_2025}. Lavender is allegedly, but not confirmed, to be no longer in use in part due to international pressure \cite{abraham_2024a,un_2024}. In contrast, all other systems are actively expanding their user base and/or geographic reach\cite{flocksafety_2025d,flocksafety_2026,douglas_2025,healy_2024,human_rights_watch_2020,office_2025,white_2025,borak_2024,borek_2023,korolev_zhabin_2024}.

Lastly, notwithstanding our findings, there are notable information gaps regarding each system’s lifecycle across the ten dimensions. The most salient information gaps across systems, regardless of political regime, were related to the design and development and desting stages.  With the exception of SlimmeCheck \cite{lighthouse-reports_2025,guo_geiger_braun_2025}, there is limited or incomplete information regarding model specifications, training data, and the technical stack each system is embedded in. With the exception of SlimmeCheck and the Live Facial Recognition (LFR) Vans, there is limited information regarding how these systems were evaluated. Notably, both systems were evaluated by third-party evaluators in collaboration with local government officials \cite{braun_geiger_guo_constantaras_silverman_2025,fussey_murray_2019,national_physical_laboratory_2020}. For systems deployed in authoritarian regimes (IJOP, Sfera) and the Lavender system used by the Israeli military, there is a greater lack of information regarding oversight mechanisms and system safeguards. 

\section{Discussion}
{Our findings extend existing scholarship on AI-enabled authoritarianism in three ways. First, we find that authoritarian-enabling features are present across political contexts, including liberal democratic ones. This suggests regime classification is insufficient for understanding authoritarian risks, and a practice-oriented approach is needed\cite{glasius_2018}. Secondly, through our analysis of FlockSafety's fragmented, privately operated ALPR system, we challenge the assumption that AI-enabled authoritarianism is strictly enacted by states in formally classified authoritarian regimes. Third, by shifting beyond model development towards the full system lifecycle, we demonstrate how authoritarian outcomes emerge from design and deployment choices that capability-focused analyses overlook.

Rather than classifying system outcomes as the product of regime type alone, our findings demonstrate that AI-enabled authoritarianism can also be enabled by design and deployment choices. Across political regimes, we uncover how specific features help enable key characteristics of authoritarianism, reframing anti-authoritarian AI governance as a wider socio-technical problem rather than a concern confined to non-democratic contexts. These features include widespread data integration, the co-optation of civil administrative systems, and weak or ineffective governance instruments. The role of regimes is not determinative, as we find systems enable authoritarian characteristics, such as the expansion of coercive capacity and the erosion of accountability mechanisms, in both liberal democracies and autocracies. This is also evident in Section 5.2, as the lifecycle features of systems do not align into clear, regime-specific patterns. Instead, there is considerable variance regarding the configuration and combination of these features across regimes.

Crucially, we note that whether systems enable authoritarianism is not restricted to a particular model, nor does it depend on highly capable models. In fact, both centralized and fragmented systems can enable authoritarianism due to their respective unique ability to evade oversight. Instead, authoritarian impacts emerge from user and developer choices embedded throughout the system lifecycle, from ideation and procurement to integration and rollout. Therefore, such impacts are not inevitable. We thus encourage policymakers and technical developers to treat AI-enabled authoritarianism as a current risk that requires mitigation, further studies, and targeted regulation, rather than an emerging risk contingent upon the direction of the capabilities of frontier models \cite{bengio_2023,davidson_finnveden_hadshar_2025}. AI-enabled authoritarianism warrants present concern, as it can occur without the development of artificial general intelligence\cite{hendrycks_song_szegedy_lee_gal_brynjolfsson_li_zou_levine_han_et_al_2025}. Future research should consider how embedded user and developer choices influence the systems’ capacity to enable authoritarianism as well, rather than narrowly focusing on technical capabilities and political deployment contexts. 

Consequently, mitigation efforts should focus not only on downstream safeguards or user compliance post-deployment, but earlier lifecycle stages as well, in which systems are justified, procured, developed, and integrated with other systems. We recommend stakeholders consider interventions across the entire system lifecycle to minimize the risk that systems contribute to AI-enabled authoritarianism. Notwithstanding these findings, we acknowledge that the recommended interventions cannot entirely eliminate AI-enabled authoritarianism on their own. Rather they function to mitigate risk within the scope of the stakeholders’ influence. \subsection{Implications for Developers}
Our findings reveal that operational safeguards alone are insufficient, as they rely on user compliance. The case study of the Lavender and FlockSafety's ALPR system illustrates how users can neglect or easily bypass operational practices. This underscores the need for technical safeguards embedded directly into systems as an additional defense against user misuse. 

Developers should assume that government and public infrastructure systems may be repurposed after deployment and should proactively design for resistance to misuse. As this will require developers to anticipate how features can be misused beyond their original intent, we re-emphasize the need for further studies on the specific feature-level mechanisms that enable authoritarianism. For example, boundary control risks can emerge when features that encode ethnicity, religion, nationality, or their proxies are intentionally exploited for discriminatory \footnote{We use "discriminatory" in the socio-political sense to refer to prejudicial distinctions that impair the recognitions, enjoyment, and exercise of rights between categories of people, particularly along lines of ethnicity, religion, nationality, and other protected characteristics} \cite{hrcgc18}, rather than in the technical machine learning sense of model differentiation between classes} purposes. Safeguards should be prioritized and subjected to the same level of rigor as any other part of the system’s specifications.  
\subsection{Implications for Policy Makers}
Transparency varies considerably between systems. For example, investigations revealed details of the IJOP system’s technical stack in China, but the same level of detail is not available for Flock Safety’s ALPR system in the US. These differences in data availability are not merely methodological constraints but likely reflect meaningful variations in oversight, governance, and overall accountability. As a result, policymakers should consider mandating increased transparency from technical developers and administrators. 
 
One critical intervention point lies in the testing stage. Establishing performance red lines before systems are piloted is crucial. Clear success criteria and evaluation data lower the barriers for decision-makers to roll back the deployment of systems and save face. For example, during the dismantling of SlimmeCheck, Amsterdam city officials cited evaluation data that demonstrated the system performed worse than the existing analogue system\cite{guo_geiger_braun_2025}. Such documentation can serve as critical empirical justification.

While developers build safeguards during system design, policymakers can enforce accountability during other stages of the system’s lifecycle, especially once systems are deployed. These stages include: procurement, testing, operational rollout, oversight, and current stage. However, enforcing accountability will require real legislative authority, with clear standards for use, misuse, and consequences. For example, to limit high degrees of centralization and integration of administrative data, policymakers should enforce strict data-sharing regulations between agencies with explicit data retention limits. These will help close regulatory gaps, such as those of the UK’s live facial recognition policies. 
\subsection{Implications for Liberal Democracies}
Progression towards authoritarian outcomes via AI systems can occur within democratic institutions and is not limited to authoritarian regimes. The erosion of democratic practices is not limited to overt repression or institutional breakdown. Instead, authoritarianism can be enabled incrementally, which can be facilitated through technology, as evidenced by our research. As liberal democracies are not immune to AI-enabled authoritarianism, we advocate for greater scrutiny of technological systems used by or in government and law enforcement officials, as the risk of AI-enabled authoritarianism may be overlooked and minimized in these contexts.  

Next, many of the systems have been adopted and justified in the name of efficiency, modernization, and public safety, as demonstrated by the Sfera, Flock Safety’s ALPR, and LFR Van systems. This suggests that there is a delicate tradeoff between these goals and accountability, so liberal democracies must be vigilant in adopting the necessary safeguards to prevent the scope creep of these AI systems. 

Lastly, our examination of FlockSafety’s ALPR system reveals that fragmented deployments can uniquely enable authoritarianism by diffusing responsibility for surveillance across stakeholders, weakening overall accountability. Liberal democracies, therefore, cannot ignore private-sector involvement and must regulate private entities that develop and deploy fragmented AI systems that exploit the nascent regulatory landscape. This finding enriches our analysis by demonstrating how AI-enabled authoritarian practices can occur outside authoritarian regimes and are not exclusively enacted by state actors.

\subsection{Limitations}
Several limitations exist that should inform the interpretation of our results. First, due to the nature of this research, there are substantial information gaps that limit a comprehensive analysis. Despite extensive efforts to collect information on these controversial systems, key technical and operational details are not fully available online for public viewing\cite{wang_2019,braun_geiger_guo_constantaras_silverman_2025,aclu_oregan_2025,nec_australia_2014}. This is due to a combination of factors, including outdated records, intentional efforts to limit transparency, and corporate secrecy that hides the internal technical workings of proprietary technologies.

As we limit this research to only six case studies, these findings cannot be widely generalized across all contexts and AI deployments. Additionally, we acknowledge limits in comparability as the six systems vary considerably in scale, purpose, technical design, and context. Our findings should thus be strictly interpreted through the lens of the authoritarian-enabling features identified in high-level system analysis. We encourage readers to view our findings as an initial contribution to a nascent intersection of research, rather than a comprehensive and complete account of AI-enabled authoritarianism.

As stated previously, many of the governance characteristics identified in our framework in Section 2.1 are not found exclusively in authoritarian regimes, nor do they manifest only along the democratic-authoritarian spectrum. For example, many of these characteristics intersect with mechanisms employed by colonial administrative structures, such as administrative data systems and population registries\cite{Breckenridge_2016,Fanon_Sartre_1963,Scott_1998,Chatterjee_1993}. This suggests the democratic-authoritarian spectrum is likely just one analytical lens for understanding AI-enabled governance harms, and a fuller account would require further engagement with non-Western and postcolonial scholarship. We invite researchers to enrich this analysis by considering additional dimensions our approach does not fully capture, and that are underexplored in the existing literature.

Future directions include expanding the number of systems investigated, political contexts, and units of comparison, considering features that enable democracy instead, and drawing on information available in other languages and platforms not examined. Additionally, in-depth studies of single-system deployments and their specific contexts can provide more granular accounts of enabling factors. Lastly, as the relationship between technology and authoritarianism continues to evolve, we encourage researchers, particularly from the political and social sciences, to further explore the specific mechanisms by which AI systems can abet or constrain anti-democratic practices, and how regime type impacts those mechanisms.  

\section{Conclusion}

This study offers one of the few systematic, cross-regime comparisons of the technical and operational features of AI systems that enable authoritarianism.As such, this study extends the current literature by demonstrating that enabling features do not map neatly to regime type, challenging assumptions about how authoritarian-enabling systems are deployed, and widening the scope of analysis beyond model development to include deployment.
Our research, based on the mapping of these systems across ten lifecycle stages, reveals that six characteristics of authoritarianism (coercive capacity, accountability erosion, symbolic safeguards, information control, anticipatory repression, and boundary control) can be furthered through recurring patterns of design and deployment choices, regardless of political regime. 

We identify these features as a) overreliance on automation, b) mass data integration and centralization, c) co-optation and integration of administrative and military infrastructures, d) oversight and regulatory gaps, e) weak user compliance, and f) harms to protected groups. Importantly, these features are not necessarily tied to highly accurate or advanced AI systems. We examine the implications and provide suggestions for developers, policy makers, and liberal democracies. We conclude by reemphasizing that AI-enabled authoritarianism should be understood as an ongoing risk that requires mitigation, attention, and governance across the full system lifecycle.
 
\section{Generative AI Usage Statement}

During the preparation of this work, the authors used Grammarly to assist with grammar and spell checking. ChatGPT was used to format tables, and check for inconsistent usage of key terms. Grammarly, ChatGPT, and Claude were used to suggest changes in sentence-level fluency. No generative AI was used to develop the core ideas or structure of the paper. 

\section{Acknowledgements}
The research was supported by the  authors' affiliation with the Oxford Martin AI Governance Initiative, funded by the Berkeley Existential Risk Initiative. We thank Keir Reid, Dr. Isaac Friend, Igor Krawczuk, Rose Hadshar, Uma Kalkar, Shariqah Hossain, Garrett Sanborn, Emma Pan and Tappy Lung  for their feedback in shaping this work. We thank Professor Darren Byler from Simon Fraser University and Conor Healy from Internet Protocol Video Market (IPVM) for their time, and invaluable expertise on surveillance systems in the US and China.

\bibliographystyle{ACM-Reference-Format}
\bibliography{sample-base}

\appendix

\begin{table*}[t]
\caption{System Mapping Against Inclusion Criteria}
\centering
\label{tab:system_inclusion_criteria}
\begin{tabular}{%
  p{0.15\textwidth}
  p{0.40\textwidth}
  p{0.20\textwidth}
  p{0.15\textwidth}
}
\toprule
\textbf{System Name} &
\textbf{Presence of Artificial Intelligence} &
\textbf{System Classification} &
\textbf{Political Regime Classification} \\
\midrule

FlockSafety Automated License Plate Recognition (ALPR) &
While the AI/ML model is not known, the system is deduced to have computer vision capabilities based on its use case. &
Surveillance &
Liberal Democracy\cite{nord_altman_angiolillo_fernandes_good_god_lindberg_2024}, Free\cite{gorokhovskaia_grothe_2025} \\

Live Facial Recognition (LFR) Vans &
The system is confirmed to be AI/ML-based, as verified by vendor and third-party testing\cite{grother_ngan_hanaoka_2026}. As of 2013, the system relied on a modified Generalized Learning Vector Quantization (GLVQ) algorithm\cite{nec_australia_2014}. &
Surveillance &
Electoral Democracy\cite{nord_altman_angiolillo_fernandes_good_god_lindberg_2024}, Free\cite{gorokhovskaia_grothe_2025} \\

SlimmeCheck &
The system is confirmed to be AI/ML-based, according to developers and third-party testers. It relies on an explainable boosting machine algorithm\cite{braun_geiger_guo_constantaras_silverman_2025}. &
Predictive Policing &
Liberal Democracy\cite{nord_altman_angiolillo_fernandes_good_god_lindberg_2024}, Free\cite{gorokhovskaia_grothe_2025} \\

Lavender &
The system is confirmed to be AI/ML-based, as indicated by its developer. The system leverages positive unlabeled learning\cite{human_rights_watch_2024,tauvod_2023}. &
Predictive Policing &
Electoral Democracy\cite{nord_altman_angiolillo_fernandes_good_god_lindberg_2024}, Free\cite{gorokhovskaia_grothe_2025} \\

Integrated Joint Operations Platform (IJOP) &
The surveillance apparatus part of the system is confirmed to be AI/ML-based on its various developers\cite{ipvm_editorial_staff_2021}. &
Surveillance, Predictive Policing &
Closed Autocracy\cite{nord_altman_angiolillo_fernandes_good_god_lindberg_2024}, Not Free\cite{gorokhovskaia_grothe_2025} \\

Sfera &
The system is confirmed to be AI/ML-based on its developers\cite{coldewey_2016,visionlabs_2026,grother_ngan_hanaoka_2026}. &
Surveillance, Predictive Policing &
Electoral Autocracy\cite{nord_altman_angiolillo_fernandes_good_god_lindberg_2024}, Not Free\cite{gorokhovskaia_grothe_2025} \\

\bottomrule
\end{tabular}
\end{table*}

\section{Appendix}
\begin{table*}[t]
\caption{Mapping of Authoritarian Characteristics to Enabling Features and Deployed Systems.}
\centering
\small
\renewcommand{\arraystretch}{1.15}
\setlength{\tabcolsep}{6pt}
\label{tab:results}

\begin{tabular}{p{0.25\textwidth} p{0.45\textwidth} p{0.20\textwidth}}
\toprule
\textbf{Authoritarian Characteristic} & \textbf{Enabling Feature} & \textbf{System} \\
\midrule

\multirow{5}{=}{\textbf{Coercive Capacity}} &
Integration with Lethal Military Infrastructure & Lavender \\
& Insufficient Accuracy Rates & Lavender \\
& Widespread Data Integration & ALPR System; IJOP; LFR Vans; Sfera \\
& Integration with Outdated Databases & ALPR System \\
& Co-optation of Administrative Systems & IJOP; LFR Vans; Sfera \\
\midrule

\multirow{4}{=}{\textbf{Accountability Erosion}} &
Inconsistent Operating Procedures & LFR Vans \\
& Overreliance on Automated Decision-Making & Lavender \\
& Oversight Loopholes for Non-state Actors and Private Citizens/Organizations & ALPR System \\
& Regulatory Gap and Exemption for Law Enforcement & Sfera \\
\midrule

\multirow{2}{=}{\textbf{Symbolic Safeguards}} &
Lack of Compliance Mechanisms & ALPR System; Lavender \\
& Inconsistent Operating Procedures & ALPR System \\
\midrule

\multirow{2}{=}{\textbf{Information Control}} &
Widespread Data Integration & IJOP \\
& Co-optation of Administrative Systems & IJOP; Sfera \\
\midrule

\multirow{3}{=}{\textbf{Anticipatory Repression}} &
Co-optation of Administrative Systems & Sfera \\
& Predictive Behavioural Indicators & IJOP \\
& Arbitrary Indicators & IJOP \\
\midrule

\multirow{4}{=}{\textbf{Boundary Control}} &
Explicit Group-Based Classification & IJOP \\
& Arbitrary Indicators & Lavender \\
& Biased Training Data & SlimmeCheck \\
& Encoding of Protected Group Characteristics & SlimmeCheck \\
\bottomrule
\end{tabular}
\end{table*}

\end{document}